\documentstyle[12pt,aps]{revtex}

\tighten

\begin{document}
\newfont{\bftitle}{cmbx12 scaled\magstep5}
\newcommand{\be}{\begin{equation}}
\newcommand{\bea}{\begin{eqnarray}}
\newcommand{\eea}{\end{eqnarray}}
\newcommand{\ee}{\end{equation}}
\newcommand{\bb}{\langle}
\newcommand{\kk}{\rangle}
\newcommand{\bk}[4]{\bb #1\,#2 \!\mid\! #3\,#4 \kk}
\newcommand{\kb}[4]{\mid\!#1\,#2 \!\mid}
\newcommand{\kx}[2]{\mid\! #1\,#2 \kk}
\def\notp{{\not\! p}}
\def\up{{\uparrow}}
\def\down{{\downarrow}}
\def\bfq {{\bf q}}
\def\bfK{{\bf K}}
\def\bfk{{\bf k}}
\def\bfp{{\bf p}}                                

\title{\hskip3in {NT@UW-01-32,   INT-PUB 02-26}\\ \hskip 1in\\
  $Q^2$ Independence of  $QF_2/F_1$, Poincare Invariance and the
  Non-Conservation  of Helicity}

\author{Gerald A. Miller}
\address{Department  of Physics, Box 351560,
University of Washington, Seattle, WA 98195-1560, USA}
\author{Michael R. Frank}
\address{Institute For Nuclear Theory, Box 351550, and Department of Physics,
University of Washington, Seattle, WA 98195-1560, USA}  
\maketitle


\begin{abstract}
A relativistic constituent quark model is found to reproduce the recent
data regarding the ratio of proton form factors, $F_2(Q^2)/F_1(Q^2)$. 
We show that imposing Poincare invariance leads to substantial violation of
the helicity conservation rule,  as well as 
an analytic result that  the ratio $F_2(Q^2)/F_1(Q^2)\sim 1/Q$ for intermediate
values of $Q^2$. 
\end{abstract}

\section{Introduction}
The recent discovery
\cite{Jones:1999rz},
\cite{Gayou:2001qd}
that the ratio of proton form factors $G_{E}/G_{M}$ falls linearly with 
$Q^2$ and   that the ratio  $QF_{2}/F_{1}$ reaches a constant value
for $2\le Q^2\le 6 $ GeV$^2$ has focused attention on understanding
nucleon structure.            The constant  nature of the
ratio  $QF_{2}/F_{1}$  contrasts with the prediction from 
perturbative QCD\cite{muellerpr,BL80}
that $Q^2F_{2}/F_{1}$ should be constant. While this latter ratio  
could be achieved when experiments are pushed to higher values
of $Q^2$, it is worthwhile to obtain a deeper understanding of the present
results. In particular,  
the perturbative result is based on the notion that helicity is
conserved\cite{Brodsky:1981kj} in high momentum 
transfer exclusive processes, so that it becomes interesting to understand
why this 
conservation does not seem to be applicable.

The qualitative nature of the experimental results were anticipated
or reproduced by several model calculations
\cite{Frank:1995pv,Holzwarth:1996xq,Cardarelli:2000tk,Wagenbrunn:2000es},
 with that of Ref.~\cite{Frank:1995pv}  
based on the work of Schlumpf\cite{fs1,fs2,fs3} being the earliest.
The implementation of relativity is an important feature of each of these
calculations, so it is natural seek an understanding of the form factors
in terms of relativity. Our purpose here is to examine the model of 
Ref.~\cite{Frank:1995pv} with the aim of highlighting the essential features
which  cause the ratio $QF_{2}/F_{1}$ to be constant.

We proceed by   presenting definitions and kinematics relevant for a light
front analysis in Sect.~II. The relevant features of our relativistic
constituent
quark model are displayed
in Sect.~III. The essential reason for the constant
ratio
$QF_2/F_1$ is discussed  in Sect.~IV, and elaborated upon numerically
in Sect.~V. The paper is concluded with  a brief summary.

\section{Definitions and Kinematics}

The electromagnetic current matrix element can be written
in terms of two form factors taking into account current and parity
conservation:

\begin{equation}
\left< N,\lambda ' p' \left| J^\mu \right| N,\lambda p\right> =
\bar u_{\lambda '}(p') \left[ F_1(Q^2)\gamma^\mu + {\kappa F_2(Q^2) \over
2 M_N}i\sigma^{\mu\nu}(p'-p)_\nu \right] u_\lambda (p)
\label{eq:3.1}
\end{equation}
with momentum transfer $q^\mu=(p'-p)^\mu$,
$Q^2=-q^2$ and $J^\mu$ is taken as the electromagnetic current of free
quarks.
For $Q^2 = 0$ the form factors $F_1$ and $\kappa F_2$ are respectively equal
to the charge and the anomalous magnetic moment $\kappa$ in  units
 $e$ and
$e/M_N$, and the magnetic moment is $\mu = F_1(0) + \kappa F_2(0)=1+\kappa$.
The Sachs form factors are defined as
\begin{equation}
G_E = F_1 - {Q^2 \over 4M_N^2}\kappa F_2\;, \quad\hbox{and}\quad
G_M = F_1 +  \kappa F_2\;,
\end{equation}

The evaluation of the form factors is simplified by using the
so-called Drell Yan reference frame in  which  $q^+=0$ so that
$Q^2=q_{\perp}^2$. This means that   the  plus components of the
nucleon momenta  (and also those of the struck constituent quark) 
are not
changed by the absorption of the incoming photon.

If light-front spinors for the nucleons are used,
the form factors can be expressed simply in terms of the plus-component of
the current\cite{bd}:
\begin{eqnarray}
F_1(Q^2) &=&{1 \over 2P^+}\left< N,\uparrow\left| J^+\right| N,
\uparrow\right>\;, \quad\hbox{and}\quad
Q\kappa  F_2(Q^2) ={-2M_N \over 2P^+}\left< N,\uparrow\left|
J^+\right| N,\downarrow\right>\;.
\label{ffs}
\end{eqnarray}
The form factors are calculated,
using 
the ``good" component of the current, $J^+$,
to suppress the effects of quark-pair terms. 

It is worthwhile to compare the formalism embodied  in Eq.~(\ref{ffs})
with the non-relativistic quark
model formalism in which $G_E$ and $G_M$ are  the Fourier transforms
of the ground state  matrix elements of the  quark charge ($\sum_{i=1,3}
e_i\delta ({\bf r}-{\bf r_i})$ 
and magnetization $\sum_{i=1,3}
{e_i\over 2m_i}\delta ({\bf r}-{\bf r_i})$ density operators. At 
high momentum transfer one needs to account for the influence of the 
motion of the proton on its wave function. Since the charged quarks of the
initial proton  state (or final state, or both initial and final states) 
are moving, charge and magnetic effects are correlated in a 
manner consistent with relativity. It is necessary
to maintain this relativistic connection, lost
 in the non-relativistic quark model.
The use of light front dynamics,  concomitant with Eq.~(\ref{ffs}),
  is a particularly convenient way to handle the motion of the initial and
final state. This is because the proton
wave function, 
a function of internal 
relative coordinates, is the same in any reference frame.
\section{Relativistic Constituent Quark Model of The nucleon} 

We study the form factors
using relativistic constituent quark models in general, and starting with
the model  of Schlumpf\cite{fs1,fs2,fs3} in particular. 
Such models have a long history\cite{bere76,bere77,bakk79}, and
many authors  
\cite{FS81,azna82,jaus90,chun91,cc,dzie88,webe87,sc,gm,ss1,keister,kone90,jaco90,hill90,deAraujo:1999hn,Bodoor:2000ic} 
have 
contributed to the necessary developments.
Schlumpf's model is used because his power-law wave   
functions lead to a reasonably good description of the 
proton electromagnetic form factors, $G_E$ and $G_M$, at all of the values of
$Q^2$ where data were available as of 1992\cite{fs1,fs2,fs3}.
This model uses the Bakamjian-Thomas BT construction, which 
implies the choice of a very specific model wave function
\cite{webe87,deAraujo:1999hn}. The use of a definite model allows us  
 to gain insight, but  we also
shall discuss the limitations of this approach.

We remind the reader about a few basic features of light front treatments in
the BT approach.
The light-front formalism is specified by the invariant
hypersurface $x^+ = x^0+x^3 =$ constant. The following notation is used: A
four-vector $A^\mu$ is given by $A^\mu = (A^+,A^-,{\bf A}_\perp)$,
where  $A^\pm \equiv A^0 \pm A^3$ and ${\bf A}_\perp=(A^1,A^2)$.
Light-front momenta vectors are denoted by  ${\bf p} =
(p^+,{\bf p}_\perp)$, with
$p^-=(p_\perp^2+m^2)/p^+$ for on-shell quarks.
The three momenta $\bf{p}_i  $ of the quarks
can be transformed to the total and relative momenta to facilitate
the separation of the center of mass motion
 as
\begin{eqnarray}
\bf {P}&=& {\bf p}_1+{\bf p}_2+ {\bf p}_3,
  \qquad \xi={p_1^+\over p_1^++p_2^+}\;,
\qquad
\eta={p_1^++p_2^+\over P^+}\;,\nonumber\\
{\bf k}_\perp &=&(1-\xi){\bf p}_{1\perp}-\xi {\bf p}_{2\perp}\;, \quad
{\bf K}_\perp =(1-\eta)({\bf p}_{1\perp}
+{\bf p}_{2\perp})-\eta {\bf p}_{3\perp}\;.
\label{relv}
\end{eqnarray}
It is also useful to consider the mass operator of a non-interacting system of
total
momentum $P^\mu$:
\bea
M_0^2\equiv\sum_{i=1,3} p^-_i\;P^+-P_\perp^2=
{K_\perp^2\over \eta(1-\eta)}+
{k_\perp^2+m^2\over \eta\xi(1-\xi)} +{m^2\over 1-\eta},\label{m0}\eea
where $m$ is the light quark mass, taken as the same for up and down quarks.

One may express the proton wave function in the center of mass  frame in which 
the individual momenta are given by
\begin{eqnarray}
{\bf p}_{1\perp}&=&{\bf k}_\perp+\xi {\bf K}_\perp,\;\;\quad
{\bf p}_{2\perp}=-{\bf k}_\perp+(1-\xi){\bf K}_\perp\;,\quad 
{\bf p}_{3\perp}=-{\bf K}_\perp.\label{pi}
\end{eqnarray}
The use of 
light front variables enables one to 
separate the center of mass motion from the
internal motion. The internal wave function $\Psi$ is 
therefore a function of the
relative momenta ${\bf p}_i,\xi,\eta$.
The internal wave function of the proton depends on these relative momenta.
One can obtain the wave function in  certain special frames via
a kinematic boost. In particular, if the proton acquires a transverse
momentum (is boosted) by the  absorption of a photon of momentum
$\bfq=(0,\bfq_{\perp})$   by the third quark,      
the effects of the boost are obtained merely
replacing the momenta $\bfk_\perp, \bfK_\perp$                  by
\be
\bfk'_\perp=\bfk_\perp,\qquad 
\bfK'_\perp=\bfK_\perp-\eta\bfq_{\perp}\label{prime}\ee
The mass operator, $ M_0'$  of the boosted system  is obtained by replacing
${\bf k}_\perp,{\bf K}_\perp$ by the variables of Eq.~(\ref{prime}) so that
\bea
{M_0'}^2=
M_0^2+{-2\eta \bfK_\perp\cdot\bfq_\perp+\eta^2q_\perp^2\over \eta(1-\eta)}.
\label{defmop}\eea

We now turn to the construction of the non-perturbative wave function,
$\Psi$. This
is based on the attempt to construct a state, described in terms of the given
light front variables, that also is an eigenstate of  angular
momentum\cite{chun91}.
We take the proton wave function  to be  a product
of an anti-symmetric color wave function  with a 
symmetric flavor-spin-momentum 
wave function $\Psi$.
To understand the construction\cite{chun91} 
of the relativistic wave function, 
it is worthwhile to 
 start by considering  
 the non-relativistic model.
Then 
\bea\Psi^{NR}={1\over\sqrt{2}}
\left(\phi_\rho\;\chi_\rho+\phi_\lambda\;\chi_\lambda\right)\Phi \label{nr}\eea
  where $\phi_{\rho}$ represents a mixed-antisymmetric and $\phi_{\lambda}$
a   mixed-symmetric flavor wave function and, $\chi_{\rho,\lambda}$
  represents mixed symmetric or anti-symmetric
 spin wave functions (in terms of Pauli spinors).
In the non-relativistic model  
the wave function $\Phi$ depends on spatial variables only, and the  
computed form factors $G_E$ and $G_M$
will have the same dependence on $Q^2$.

 The relativistic
  generalization of Eq.~(\ref{nr}) is
  \bea
\Psi(p_i)&=&u(p_1) u(p_2) u(p_3)\psi(p_1,p_2,p_3),\label{psieq}\eea
where  $p_i$ represents space, spin and isospin indices:
$p_i=\bfp_i  s_i,\tau_i$ and repeated indices are
summed over.
The spinors $u$ are canonical Dirac spinors:
\bea u(p,s)={\notp +m\over \sqrt{E(p)+m}}\left(\begin{array}{c}
\chi^{\rm Pauli}_{s} \\
0
\end{array}
\right),  \
\label{dirac}\eea
with the isospin label suppressed.

The completely symmetric nature of the space-spin-flavor wave
function is preserved by using\cite{chun91}
\bea
\psi(p_1,p_2,p_3)&=&\Phi
\left[\bar u(p_1)\Gamma\bar u_2^T(p_2)\;\bar u_3(p_3)u_N(0)
  +\bar u(p_1)\Gamma^{\mu,\alpha}\bar u_2^T(p_2)\;\bar u_3(p_3)
  \tilde\Gamma^{\nu,\alpha}
 u_N(0) g_{\mu\nu}\right]\label{waveq}\\
\Gamma&\equiv& -{1\over\sqrt{2}}{1+\beta\over 2}\gamma_5C i\tau_2,\quad
\Gamma^{\mu,\alpha}\equiv{1\over\sqrt{6}}\gamma^\mu Ci\tau_\alpha\tau_2,\quad
\tilde\Gamma^{\nu,\alpha}\equiv{1+\beta\over 2}\gamma^\nu\gamma_5\tau_\alpha,
\eea
where the charge-conjugation matrix $C\equiv
i\gamma^2\gamma^0=-i\gamma_5\sigma_2$. 
Note that if one takes the non-relativistic limit of (\ref{waveq}) by taking
${\bf p}_i\to (m,{\bf 0}_\perp)$ one gets  (\ref{nr}) for the spin-isospin
dependence.

The momentum wave function 
can be chosen as a function of $M_0$ to fulfill the requirements of 
spherical and
permutation symmetry. We take the 
 $S$-state orbital function $\Phi(M)$ to be of a  power law form:
\begin{equation}
\Phi(M_0)={N\over (M^2_0+\beta^2)^{\gamma}}\;,
\label{eq:2.34}
\end{equation}
which depends on two free parameters, the constituent quark mass and the
confinement scale parameter $\beta$. 
 Schlumpf's
 parameters are  
$\beta$=0.607 GeV, $\gamma=3.5$, 
and the constituent quark 
mass, $m$=0.267 GeV.

The wave function of  
Eq.~(\ref{waveq}) now specified. It is an eigenstate of the angular
momentum operator (if interactions present in that operator are neglected)
and also  corresponds to  a state of vanishing orbital angular momentum. 
This state is Poincare invariant according to Ref.~\cite{chun91},
but this invariance may      be incomplete\cite{Bodoor:2000ic}.

\section{The Essential Effect}
The form factors of Eq.~(\ref{ffs})
are obtained by computing      
 the matrix elements of the current operator $J^+$. Here we take the
quarks to be elementary particles, so that the operator $J^+$ is essentially
the operator $\gamma^+$ times the charge of the quarks. 
The calculations
 may 
be simplified by making a unitary transformation which replaces
the  Dirac spinors
of Eq.~(\ref{psieq})
by light front spinors
\begin{eqnarray}
u_L(p,\lambda) =\frac{\rlap\slash p+ m}{\sqrt{2p^+}}\gamma^+
\left(\begin{array}{c}
\chi^{\rm Pauli}_{\lambda} \\0\end{array}
\right)
\ ,
\label{lf}
\end{eqnarray}
because\cite{BL80}
\bea \bar u_L(p^+,\bfp',\lambda')\gamma^+u_L(p^+,\bfp ,\lambda)=
2\delta_{\lambda\lambda'}p^+. \eea One then uses the completeness relation,
$1= \sum_\lambda u_L(p,\lambda)\bar u_L(p,\lambda)/2m,$ 
in Eq.~(\ref{psieq}), to
obtain  the light front representation for the
wave function:
  \bea
\Psi(p_i)&=&u_L(p_1,\lambda_1) u_L(p_2,\lambda_2) u_L(p_3,\lambda_3)
\psi_L(p_i,\lambda_i),\\
\psi_L(p_i,\lambda_i)&\equiv&
\langle \lambda_1\vert {\cal R}_M^\dagger(\bfp_1)\vert s_1\rangle
\langle \lambda_2\vert {\cal R}_M^\dagger(\bfp_2)\vert s_2\rangle
\langle \lambda_3\vert {\cal R}_M^\dagger(\bfp_3)\vert s_3\rangle
\psi(p_1,p_2,p_3),\label{psieq1}\eea
where ${\cal R}_M$ is a Melosh rotation\cite{melo74}
acting between Pauli spinors. For example, 
\begin{equation}
  \langle \lambda_3\vert 
{\cal R}_M^\dagger(\bfp_3)\vert s _3\rangle= \bar u_L(\bfp_3,\lambda_3)
u(\bfp_3, s_3)= \langle \lambda_3\vert
\left[ {m+(1-\eta)M_0+i{\bbox \sigma}\cdot({\bf n}\times {\bf p}_3)\over
\sqrt{(m+(1-\eta)M_0)^2+p_{3\perp}^2}}\right]\vert s_3\rangle.
\label{r3}
\end{equation}

The net result of this is that the relativistic spin effect is to replace
the Pauli spinors of Eq.~(\ref{nr}) with Melosh rotation operators acting
on the very same Pauli spinors\cite{chun91}. 
The spin-wave function of the $i$th quark is given by
\begin{equation}
\vert\uparrow \bfp_i\rangle\equiv{\cal R}_M^\dagger(\bfp_i)
\pmatrix{1\cr 0} \hbox{  and  }
\vert\uparrow \bfp_i\rangle\equiv {\cal R}_M^\dagger(\bfp_i)
\pmatrix{0\cr 1}\;.
\label{real}
\end{equation}
This means that, for example, 
 the spin wave function $\chi_\rho$ is replaced by  
a momentum-dependent spin wave function
$\vert\chi_\rho^{\rm rel}\bfp_i\rangle$
\bea\vert \chi_\rho^{\rm rel}\bfp_i\rangle
={1\over \sqrt{2}}\vert\up\bfp_1\down\bfp_2
-\down\bfp_1\up\bfp_2\rangle
\vert\up\bfp_3\rangle\equiv
\vert\chi^{\rm  rel}_0(\bfp_1,\bfp_2)\rangle
\vert\up\bfp_3\rangle \eea
for a spin +1/2  proton.  The term $\chi_0^{\rm rel}$ represents
the relativistic
generalization  spin-0 wave function of the quark pair labeled
by (1,2). Note that the momenta $\bfp_i$ are to be expressed in terms
of the relative variables $\bfk_\perp,\bfK_\perp,\bfk'_\perp,\bfK'_\perp$
of Eqs.(\ref{relv}) and (\ref{prime}). 
We shall see that the important
relativistic effect is contained in the difference between the spinors of
Eq.~(\ref{real}) and Pauli spinors.

The  next step is to simplify the 
calculation by using the symmetry of the wave function under interchange of 
particle labels 
to replace the quark current operator by three times that operator
acting only on  the third quark. The average charge of the third quark of the
mixed-symmetric flavor
wave function vanishes. 
This means that the second term of Eq.~(\ref{waveq}) 
does not contribute to proton electromagnetic form factors. 
The first term involves a mixed-antisymmetric wave function, so the
third quark carries the spin of the proton, $s_3$ of 
Eq.~(\ref{r3}). However, the helicity of
the light front spinor $\lambda_3$ can either be the same as
or different from $s_3$. The weighting of the terms is determined
by the two terms of the Melosh transformation of Eq.~(\ref{r3}).
Considering the arguments of  the final state wave function allows 
one to understand that the two terms are comparable.
The effect of the boost is incorporated 
simply by using Eq.~(\ref{prime}) in the argument of
 the final state wave function. Thus the large momentum $Q$
is involved and both terms of the Melosh rotation (\ref{r3}) are comparable.

There is clearly a substantial amplitude for a spin-up valence quark    
 to carry a negative light front helicity. No suppression of light-front
helicity-flip
 can
be expected from the use of such a wave function. This means that helicity
conservation does not occur. This conclusion has been obtained by Ralston
and collaborators\cite{ralston}, 
 based    on the presence of non-zero orbital angular momentum.
Here we have no orbital angular momentum, and
the mixture of light-front helicity
we obtain occurs 
as a result of imposing Poincare invariance on the constituent quark model.
Nonetheless, we certainly support their statements\cite{ralston}
that helicity non-conservation is an important effect.
We also note that Braun et al.\cite{Braun:2001tj} 
argue that soft non-factorizable terms in the wave function, with helicity
structure similar to ours, 
 are important at
intermediate values of $Q^2$. In our model, 
 there is no  basis for expecting light-front
helicity conservation because
the 
non-perturbative wave function is a mixture of different
light-front-helicity states.

If the value of $Q^2$ becomes  asymptotically large, the effects of the 
non-perturbative wave function may disappear and perturbative effects,
which do respect helicity conservation,  could take
over.  But for the present, we must take helicity non-conservation as a 
given and it will be worthwhile to consider the implications of this
 non-conservation.

\section{Proton Form Factors} 
We obtain the  form factors 
by using the wave function of Eq.~(\ref{psieq1})  in Eq.~(\ref {ffs}). Our
result can be expressed as
\begin{eqnarray}
F_1(Q^2) &=& 
\int\! {d^2\!q_\perp d\xi\over \xi(1-\xi)}{ d^2K_\perp d\eta\over\eta(1-\eta)}\;
\tilde\Phi^\dagger(M_0')\tilde\Phi(M_0)\;
\langle \chi^{\rm rel}_0(\bfp_1',\bfp_2')\vert
\chi^{\rm rel}_0(\bfp_1,\bfp_2)\rangle \;\langle\up\bfp_3'\vert
\up(\bfp_3)\rangle \label{f1}\\
Q\kappa F_2(Q^2) &=& 2M_N
\int\! {d^2\!q_\perp d\xi\over \xi(1-\xi)} 
{d^2K_\perp d\eta\over\eta(1-\eta)}\;
\tilde\Phi^\dagger(M'_0)\tilde\Phi(M_0)\;
\langle \chi^{\rm rel}_0(\bfp_1',\bfp_2'))\vert
\chi^{\rm rel}_0(\bfp_1,\bfp_2)\rangle\;
\langle\up\bfp_3'\vert
\down(\bfp_3)\rangle \label{f2}
\end{eqnarray}
The value of $M_0'$ is obtained by using Eq.~(\ref{defmop}),
and 
\bea \tilde \Phi(M_0)&\equiv& \sqrt{E_3E_{12}E_1\over M_0}\Phi(M_0)\\
E_1&=& \sqrt{k_\perp^2+m^2\over 4\xi(1-\xi)},\;
E_{12}
={K_\perp^2+4E_1^2+\eta^2M_0^2\over 2\eta M_0},\; E_3= M_0-E_{12}.
\eea

Numerical evaluations of these equations for large values of $Q^2$ were
presented in Ref.~\cite{Frank:1995pv}. Our aim here is to understand
the essential features of  the different $Q^2$-dependence of $F_1$ and $F_2$.
The expressions for the form factors 
differ only by the 
presence of the last factor
$\langle\up\bfp_3'\vert\down\bfp_3\rangle$, or  
$\langle\up\bfp_3'\vert\up\bfp_3\rangle,$
with $F_1$ depending on the  non-spin-flip term, and $F_2$ on the spin-flip
term\cite{other}.  
 To proceed we evaluate these overlaps, using the Melosh transformations with
$\bfp_{3\perp}=-\bfK_\perp$, and $\bfp'_{3\perp}=-\bfK_\perp+\eta\bfq_\perp$.
One computes the product of the matrices ${\cal R}_M(\bfp'_3)
{\cal R}_M^\dagger(\bfp_3)$. The upper diagonal element (non-spin-flip term)
appears in the expression for $F_1$, and the upper off-diagonal element
(spin-flip term proportional to $\bbox{\sigma}$)
determines $F_2$. The evaluation is simplified by
realizing that integration over  $d^2K_\perp$ causes  terms linear in
the component of $\bfK$ which are perpendicular to $\bfq$ to vanish.

To be definite, we take $\bfq_\perp$ to lie along the $x$-direction, 
so that we find
\bea
&&\langle\up\bfp_3'\vert\up\bfp_3\rangle= {
\left[(m+(1-\eta) M_0)(m+(1-\eta) M_0') +K_\perp^2 -\eta Q K_x\right]
\over \nonumber\sqrt{\left((m+(1-\eta)M_0)^2+\bfK_\perp^2\right)
\left((m+(1-\eta)M_0')^2+(\bfK_\perp-\eta\bfq_\perp)^2\right)}}\\
&&\langle\up\bfp_3'\vert\down\bfp_3\rangle={
\left[\eta Q (m+(1-\eta) M_0 )+(1-\eta)(M_0'-M_0) K_x
\right]\over\sqrt{\left((m+(1-\eta)M_0)^2+\bfK_\perp^2\right)
\left((m+(1-\eta)M_0')^2+(\bfK_\perp-\eta\bfq_\perp)^2\right)}} . \label{there}
\eea

We may understand the qualitative nature of               the ratio
$QF_2(Q^2)\over F_1(Q^2)$ using
the notion that the value of $Q=\sqrt{\bfq_\perp^2}$
can be much larger than the typical
       momenta, of order  $\beta=560$ MeV, which 
     appear in the wave function. Then
for $Q\gg \beta$, we may  approximate Eq.~(\ref{defmop})
by
\be M_0'\approx Q\sqrt{\eta\over 1-\eta},\label{lead}\ee
and take the terms of the bracketed expressions
of Eq.~(\ref{there}) which are proportional to $Q$ as  dominant.
Then, using Eqs.~(\ref{there}) and (\ref{lead}) in Eqs.~(\ref{f1},\ref{f2}),
we see that
each of 
$F_1$ and $QF_2$  contains an explicit factor of  $Q$,  and 
\bea 
{Q\kappa F_2^{As}\over F_1^{As}}
\approx2M_N {\langle\eta( m+(1-\eta)M_0 )+\sqrt{\eta(1-\eta)}K_x\rangle 
\over\langle -K_x+( m+(1-\eta)M_0 )\sqrt{\eta/(1-\eta)} \rangle},\label{approx}
\label{qed}\eea
where the expectation value symbols
abbreviate the  operation of multiplying by the remaining factors of
Eqs.~(\ref{f1},\ref{f2}) (without approximation) and performing the necessary
six-dimensional integral.
The terms $\eta,M_0,K_x$ can be anticipated to
 have an expectation value independent of
$Q^2$, so the 
ratio is anticipated to be  constant. 

The exact model calculation and the approximation (\ref{approx})
are compared in Fig.~1. Eq.~((\ref{approx})
qualitatively reproduces the constant nature of the ratio and its value.
 Thus
 the constant nature of the
ratio is understood  from the properties of the Melosh transformation, which
here
embodies the relativistic effects.
\begin{figure}
\unitlength1cm
\begin{picture}(11,9)(0,-10.5)
\includegraphics{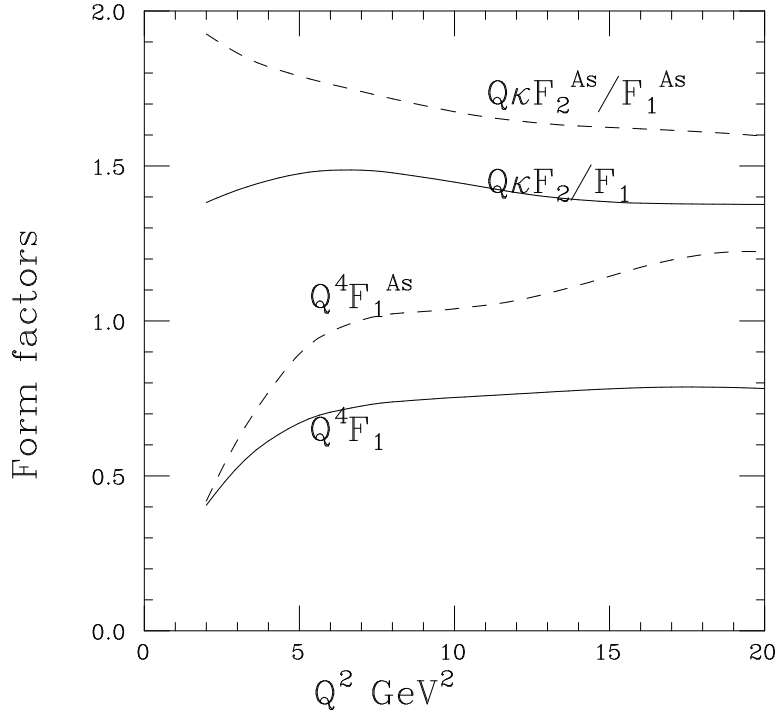}
\end{picture}
\caption{Model calculation of Eqs.~(\ref{f1},\ref{f2},\ref{there}) solid, vs. 
the approximation Eq.~(\ref{approx}), dashed.}
\label{fig:approx}
\end{figure}

Equation~(\ref{qed}) represents a
simple quick argument which gives a
constant ratio. But this  is only a rough  approximation because 
 each of $F_{1,2}$ is over predicted by about 40\%.
 Numerical work shows that neglecting the  terms proportional to
$K_x$ in both the numerator and denominator of Eq.~(\ref{approx})
leads to a different approximation:
\bea 
{Q\kappa F_2^{As}\over F_1^{As}}
\approx {\langle\eta( m+(1-\eta)M_0 )\rangle 
\over\langle ( m+(1-\eta)M_0 )\sqrt{\eta(1-\eta)} \rangle}.
\label{qed1}\eea
which, as shown  in Fig.~2
 leads to an even better reproduction of the model results for 
 $F_1$ and $F_2$.
\begin{figure}
\unitlength1cm
\begin{picture}(11,9)(0,-10.5)
\includegraphics{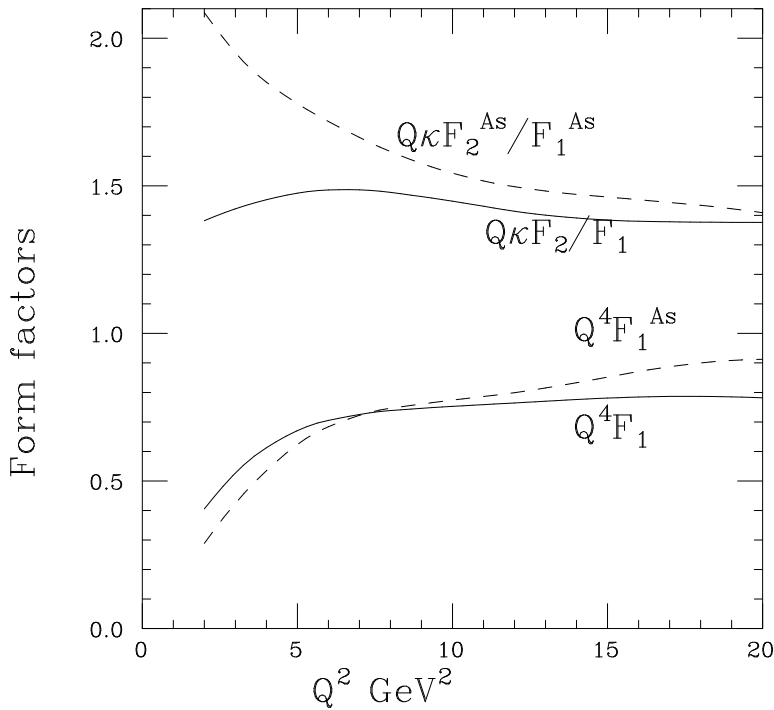}
\end{picture}
\caption{Model calculation of Eqs.~(\ref{f1},\ref{f2},\ref{there}) solid, vs. 
the approximation Eq.~(\ref{qed1}),dashed.}
\label{fig:approxqed1}
\end{figure}
Equation~(\ref{qed1})
is a better approximation because the terms involving $K_x$ cancel 
against terms involving the difference between $M_0'$ and its approximation
(\ref{lead}). Thus it seems that values of 
$Q^2$ going up to 20 GeV$^2$  are  not large enough to allow one to 
completely neglect
other terms, and therefore also not large enough to extract the asymptotic
behavior.

\begin{figure}
\unitlength1cm
\begin{picture}(11,9)(0,-10.5)
\includegraphics{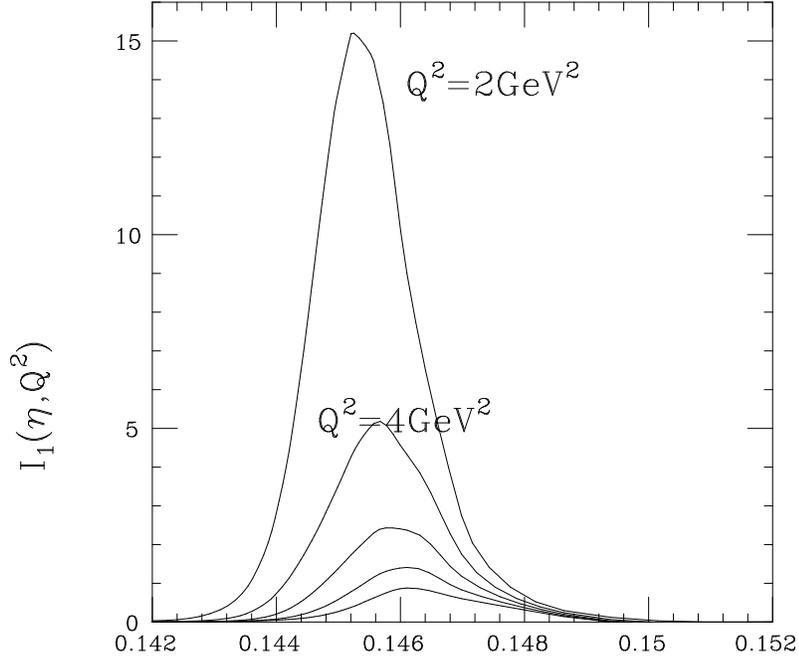}
\end{picture}
\caption{Important region of integration for $F_1$,
Eq.~(\ref{idef}). The curves show the derivative of  $I_1$
 for values of $Q^2=2,4,6,8,10$ 
GeV$^2$, with the larger values occurring for the smaller values of
$Q^2$. }
\label{fig:eta1}
\end{figure}

\begin{figure}
\unitlength1cm
\begin{picture}(11,9)(0,-10.5)
\includegraphics{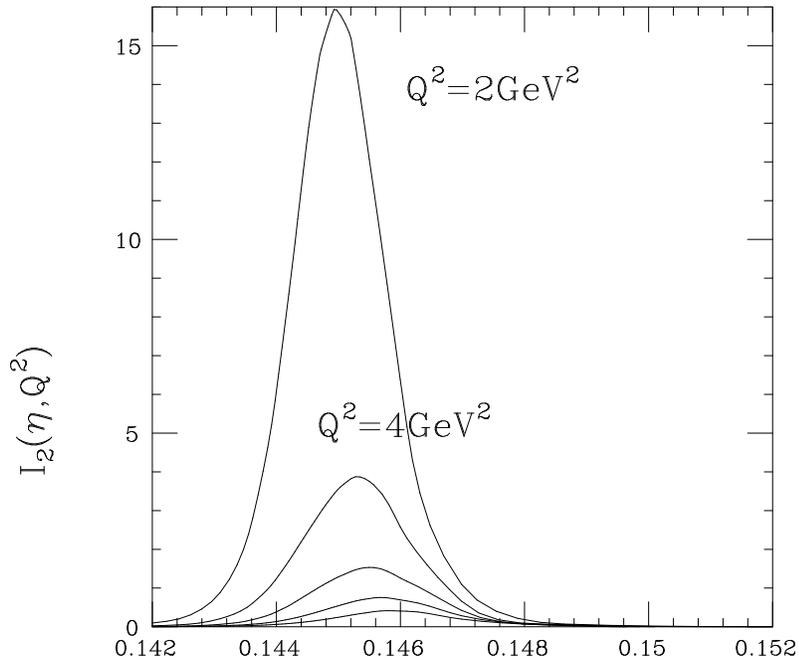}
\end{picture}
\caption{Important region of integration for $F_2$,
Eq.~(\ref{idef}) The curves show the derivative of  $I_1$
 for values of $Q^2=2,4,6,8,10$ 
GeV$^2$, with the larger values occurring for the smaller values of
$Q^2$. }
\label{fig:eta2}
\end{figure}

This feature of not reaching asymptotic values of $Q^2$
may  be understood by examining the dependence of the integrands
of Eqs.~(\ref{f1},\ref{f2}) on the value of $\eta$. We may write
\bea
F_{1,2}(Q^2)&=&\int_0^1\; d\eta\; I_{1,2}(\eta ,Q^2),\label{idef}\eea
and determine the important regions by examining $I_{1,2}(\eta,Q^2)$.
As shown in Figs.~(\ref{fig:eta1},\ref{fig:eta2}) the important contributions
occur for a very narrow band of values close to 
 $1-\eta=x_3=0.145$.  The sharp peaking is maintained
for all of the values of $Q^2$ considered here, and is a central reason
for the qualitative success of the approximations (\ref{approx},\ref{qed1}).
The small factor  $1-\eta$ multiplies  the large factor $Q$ appearing
in Eq.~(\ref{lead}), and suppresses the dominance of the terms
proportional to $Q$. The integrands peak at $x_3=0.85$, a large value
(compared to 0.33, expected if each quark were to carry the same momentum)
which
indicates the presence of the Feynman mechanism, and a corresponding
difficulty of using
simple arguments to extract
asymptotic properties of form factors.

\begin{figure}
\unitlength1cm
\begin{picture}(11,9)(0,-9.0)
\includegraphics{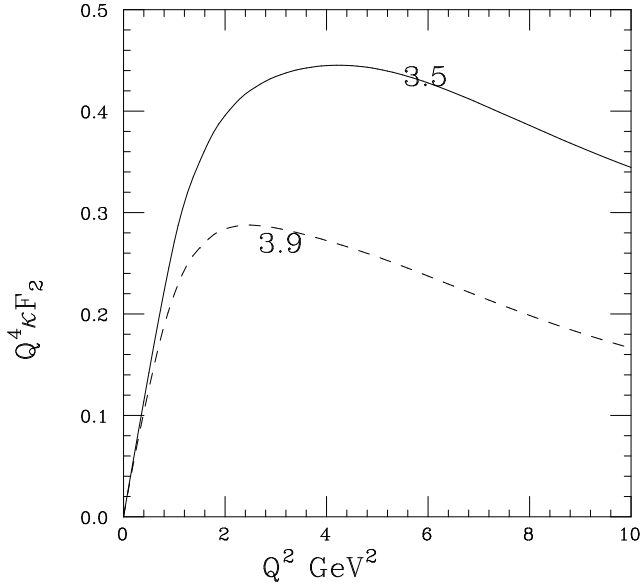}
\end{picture}
\caption{The effect of varying the parameter $\gamma$ 
which governs the power of the falloff of
the wave function of Eq.~(\ref{eq:2.34}). 
The curves for $F_2$ are labeled by the
value of $\gamma=3.5$ correct model value (solid) 
or $\gamma=3.9$ (dashed).}
\label{fig:gammaf2}
\end{figure}

\begin{figure}
\unitlength1cm
\begin{picture}(11,9)(0,-10.5)
\includegraphics{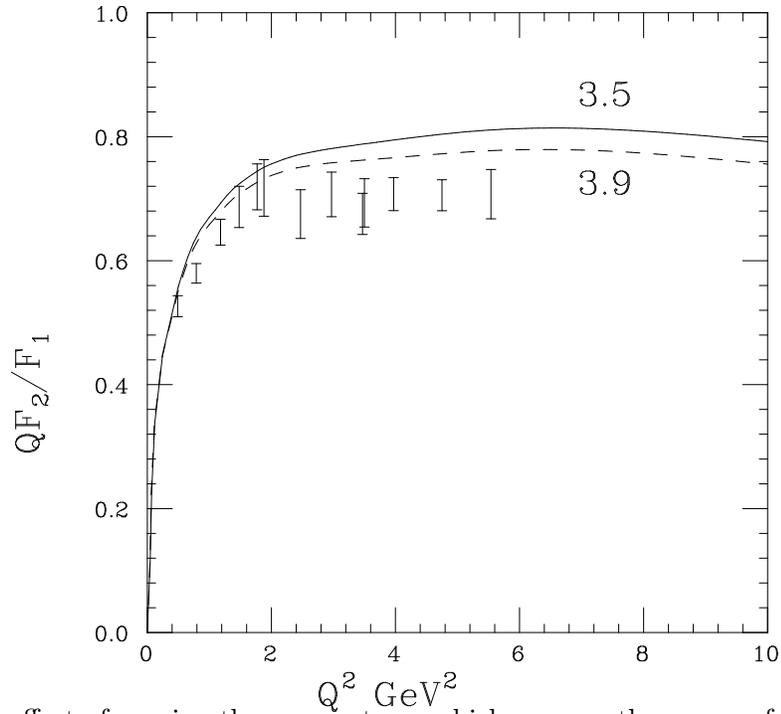}
\end{picture}
\caption{The effect of varying the parameter $\gamma$ 
which governs the power of the falloff of
the wave function of Eq.~(\ref{eq:2.34}). 
The curves for $Q\;F_2/F_1$ are labeled by the
value of $\gamma=3.5$ (solid)
or $\gamma=3.9$ (dashed).The data for $2\le Q^2\le 3.5$ GeV$^2$ are from
Ref.~1., and that for $3.5\le Q^2\le 5.5 {\rm GeV}^2$ are 
from Ref.~2.}
\label{fig:ratio}
\end{figure}

 We also find  that the computed value of the ratio $Q\kappa F_2/F_1$ is
remarkably independent of the parameters of the model. For example, 
Fig.~(\ref{fig:gammaf2}) shows that a 10\%  increase in the value of 
$\gamma$, Eq.~(\ref{eq:2.34}), causes about a 50\% decrease in 
the computed values of $F_2$,  but Fig.~(\ref{fig:ratio}) shows only  a 5\%  
 change in the
ratio. 

The  solid curve of Fig.~(\ref{fig:ratio}) represents our final results and
predictions for values of $Q^2$ that are not yet measured. The dashed curve of
Fig.~(\ref{fig:ratio}) is closer to the data for the ratio, but it corresponds
values of $F_2$ at high $Q^2$ which are much smaller than the data of 
\cite{arnold}. Thus the original model of Schlumpf seems to account for both
$F_1$ and $F_2$. There is a small disagreement ($\sim$ 15\%)
with the data for the ratio,
which can not be fixed simply by varying the parameters. This small degree of
disagreement between the model and the data is
 remarkable because so many plausible effects, such as configuration mixing 
involving both quark and gluon degrees of freedom and a non-perturbative
$Q^2$   variation of the constituent quark masses\cite{qvm}
  are ignored. Pion cloud effects\cite{cbm}
 are surely
present, but these  do not seem to be significant for  values of $Q^2$
greater than about 2 GeV$^2$.

We also study  the behavior of the ratio
$Q\kappa F_2/F_1$ for very large values of $Q^2$; see Fig.~7. The constant
nature of the ratio seen in previous figures is actually the result of
a broad maximum occurring near $Q^2\approx 10\;{\rm GeV}^2$.
The ratio  falls for 
 asymptotic values of $Q^2$, the displayed curve can
 be represented by but not as   predicted using
 perturbative QCD, which would be   $Q^2\kappa F_2/F_1\sim1/\log(Q^2)$ for
 $80<Q^2< 600 {\rm GeV}^2$.

\begin{figure}
\unitlength1cm
\begin{picture}(11,9)(0,-10.5)
\includegraphics{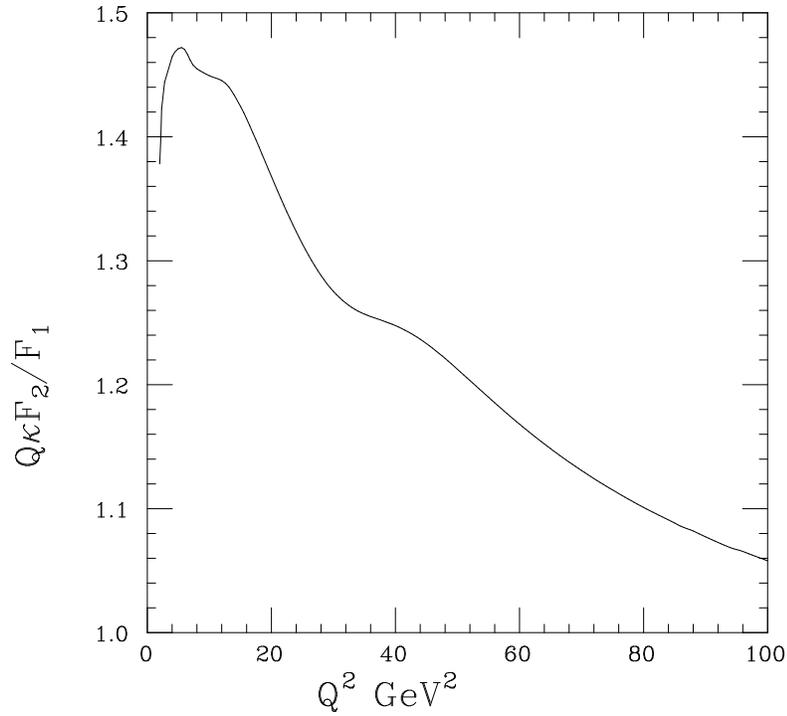}
\end{picture}
\caption{High $Q^2$ behavior  of $Q\kappa F_2/F_1$} 
\label{fig:highq}
\end{figure}

\section{Summary and Discussion}
We have seen how a simple relativistic constituent quark model 
accounts for both $F_2$ and $QF_2/F_1$ for values of $Q^2$ between 
2 and 5.5 GeV$^2$. The most relevant ingredient in the model is its
attempt to use a wave function which is Poincare invariant.
 In such 
 wave functions the helicity conservation rule is not satisfied 
 because
the 
non-perturbative wave function is a mixture of different helicity states. 
This feature leads to an analytic understanding, 
embodied in Eqs.~(\ref{approx})
and (\ref{qed1}), that $QF_2(Q^2)$ and $F_1(Q^2)$ have the same variation with
$Q^2$. The predicted value of the ratio $QF_2(Q^2)/F_1(Q^2)\approx 0.8$
for the values of $Q^2$ up to 20 GeV$^2$, and drops slowly
for larger values.  

The present data set extends to $Q^2=5.5 $ GeV$^2$, and its measurement of the
non-conservation of helicity has implications for other 
exclusive processes involving
protons. Helicity conservation should be not relevant
if  we consider proton-proton scattering at high momentum transfers, up
to $-t=5.5 $  GeV$^2$. This means large values of various analyzing powers can
be expected. Perhaps 
the most interesting mystery in proton-proton scattering is the 
large value of $A_{NN}$ observed in 90$^\circ$ proton scattering at
 $s\approx 20$
 GeV$^2$\cite{cosbie,fk}
. This corresponds to $-t\approx 7-10$ GeV$^2$. If the present measurements
are extended to values of $Q^2$ such as these, and if the constant nature
of the ratio $QF_2(Q^2)/F_1(Q^2)$ is maintained, one could be able to seek an 
explanation of the large value of the analyzing power in terms of the 
non-perturbative proton wave function.

\acknowledgements

This work was supported by the 
Department of Energy under Grant No. DE-FG03-97ER41014.
  We wish to acknowledge helpful conversations with 
S.J. Brodsky, T. Frederico,  D.S. Hwang and H.J. Weber, and to thank
O. Gayou for sending us the experimental data.



\begin{references}
\bibitem{other} Numerical evaluation shows that
 replacing the factor $ \vert\chi^{\rm  rel}_0(\bfp_1,\bfp_2)\rangle$ 
by unity
 in both the initial and final states leads to a negligible effect on the
 computed form factors for $Q^2<20$ GeV$^2$.
\end{references}
\end{document}